\documentclass{svproc}
\usepackage{graphicx}
\usepackage{url}

\begin{document}
\mainmatter              
\title{
The $S_{E1}$ factor of radiative $\alpha$ capture on $^{12}$C 
in effective field theory
}
\titlerunning{
The $S_{E1}$ factor of radiative $\alpha$ capture on $^{12}$C 
in EFT 
}  
%
\author{Shung-Ichi Ando}
\authorrunning{
S.-I. Ando
} 
\institute{
Sunmoon University, Asan, Chungman 31460, South Korea, \\
\email{
sando@sunmoon.ac.kr
} 
}

\maketitle              

\begin{abstract}
The $S_{E1}$ factor of radiative $\alpha$ capture on $^{12}$C is studied
in effective field theory. 
We briefly discuss the strategy for the calculation of the reaction 
and report a first result of $S_{E1}$ at the Gamow-peak energy,
$E_G=0.3$~MeV.
\keywords{Radiative alpha capture on carbon-12, $S_{E1}$ factor, 
effective field theory}
\end{abstract}
\section{Introduction}

The radiative $\alpha$ capture on carbon-12,
$^{12}C(\alpha,\gamma)^{16}$O, is a fundamental
reaction in nuclear-astrophysics, 
which determines the $C/O$ ratio created 
in the stars~\cite{f-rmp84}.
The reaction rate, equivalently the astrophysical $S$-factor,
of the process at the Gamow peak energy,
$E_G=0.3$~MeV, however, cannot be determined in experiment
due to the Coulomb barrier.
A theoretical model is necessary to employ in order to extrapolate
the reaction rate down to $E_G$ by fitting model parameters
to experimental data typically measured at a few MeV.
In constructing a model for the study, 
one needs to take account of
excited states of ${}^{16}$O~\cite{bb-npa06},
particularly, two excited bound states 
for $l^\pi_{i\mbox{-}th}=1_1^-$ and $2_1^+$
just below the $\alpha$-${}^{12}$C breakup threshold at
$E=-0.045$ and $-0.24$~MeV~\footnote{
The energy $E$ denotes that of the $\alpha$-${}^{12}$C system
in center of mass frame.
}, respectively,
as well as two resonant (second excited) $1_2^-$ and $2_2^+$ states at
$E=2.42$ and $2.68$~MeV, respectively.
The capture reaction to the ground state of ${}^{16}$O
at $E_G$ is expected to be $E1$ and $E2$ transition dominant
due to the subthreshold $1_1^-$ and $2_1^+$ states.
See Refs.~\cite{bb-npa06,detal-17} for review.

Theoretical frameworks employed for the previous studies
are mainly categorized into two~\cite{detal-17}:
the cluster models using generalized coordinate method~\cite{dbh-npa84}
or potential model~\cite{lk-npa85}
and the phenomenological models using the parameterization
of Breit-Wigner, $R$-matrix~\cite{lt-rmp58},
or $K$-matrix~\cite{hdz-npa76}.
A recent trend of the study is to rely on intensive numerical analysis,
in which a large amount of the experimental data relevant
to the study are accumulated, and a significant number of parameters of
the models are fitted to the data
by using computational power~\cite{detal-17,xetal-npa13,aetal-prc15}.
In the present work, 
we discuss an alternative approach to estimate the $S$-factor
at $E_G$; we employ a new method for the study
and briefly discuss a calculation of the $S_{E1}$ factor at $E_G$
based on an effective field theory~\cite{bv-arnps02,bh-pr06}.

\section{Diagrams}
\begin{figure}[t]
\begin{center}
\includegraphics[width=8.5cm]{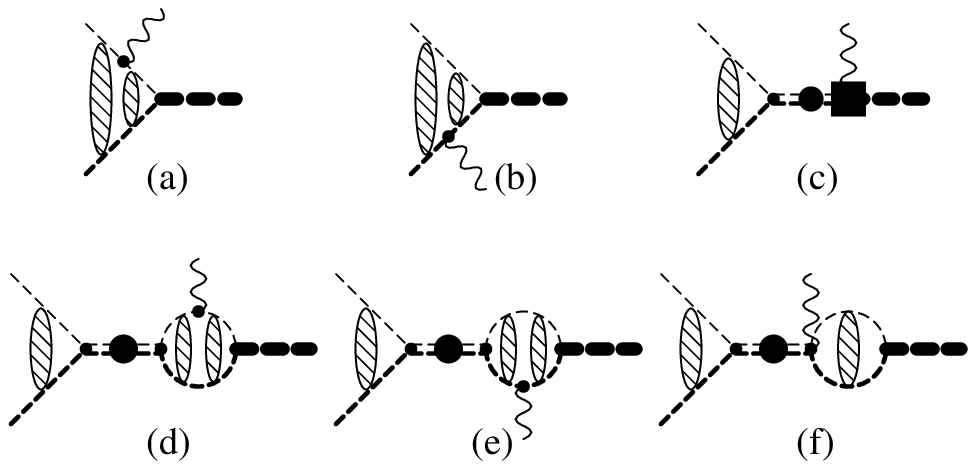}
\caption{
Diagrams for the radiative capture process from the initial $p$-wave
$\alpha$-$^{12}$C state.
A wavy line denotes the outgoing photon, and 
the same part of the diagram displayed in Fig.~\ref{fig;propagator}
is the dressed composite propagator for $l=1$ state.
A thick dashed line  
in the final state denotes the ground ($0_1^+$) state of $^{16}$O. 
See the caption of Fig.~\ref{fig;propagator} as well.
\label{fig;e1-amplitudes}
}
\vskip 4mm
\includegraphics[width=8.5cm]{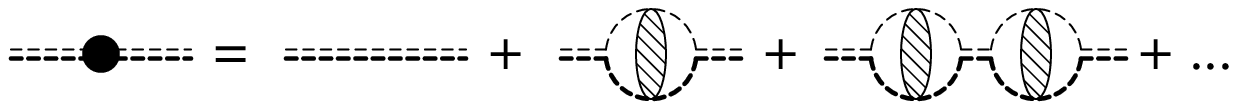}
\caption{
Diagrams for dressed $^{16}$O propagator.
A thick (thin) dashed line represents a propagator of $^{12}$C ($\alpha$),
and a thick and thin double dashed line with and without a filled blob
represent a dressed and bare $^{16}$O propagator, respectively.
A shaded blob represents a set of diagrams
for non-perturbative Coulomb interaction.
\label{fig;propagator}
}
\end{center}
\end{figure}

In the study of the radiative capture process,
$^{12}$C($\alpha$,$\gamma$)$^{16}$O, at $E_G=0.3$~MeV
employing an EFT, 
one may regard the ground states of $\alpha$ and $^{12}$C
as point-like particles
whereas the first excited states of $\alpha$ and $^{12}$C
are chosen as irrelevant degrees of freedom, from which
a large scale of the theory is determined~\cite{a-epja16}.
Thus the expansion parameter of the theory
is $Q/\Lambda_H \sim 1/3$ where $Q$ denotes a typical
momentum scale $Q\sim k_G$; $k_G$ is the Gamow peak momentum,
$k_G = \sqrt{2\mu E_G}\simeq 41$~MeV, where
$\mu$ is the reduced mass of $\alpha$ and $^{12}$C.
$\Lambda_H$ denotes a large momentum scale
$\Lambda_H \simeq 150$~MeV
obtained from the first excited energy
of $\alpha$ or $^{12}$C. 
An effective Lagrangian for the study is obtained in Eq.~(1) 
in Ref.~\cite{a-18}.

The capture amplitudes are calculated from the Feynman diagrams
depicted in Figs.~\ref{fig;e1-amplitudes} and \ref{fig;propagator}.
One can find an expression of the amplitudes in Eqs.~(6), (7), (8), and (9)
in Ref.~\cite{a-18}. We note that the loop diagrams (a) and (b) 
in Fig.~\ref{fig;e1-amplitudes} are finite whereas those (d), (e), and (f)
diverge. The divergence terms are renormalized by a counter term $h^{(1)}$ 
in a contact vertex in the diagram (c). 
Six parameters remain in the amplitudes. 
Four of them are effective range
parameters of elastic $\alpha$-$^{12}$C scattering for $l=1$~\cite{a-prc18}. 
One of them is fixed by using the binding energy of the subthreshold 
$l=1$ state of $^{16}$O,
and the others are fitted to the phase shift data 
of the elastic scattering~\cite{tetal-prc09}.

\section{Result}

\begin{figure}[t]
\begin{center}
\includegraphics[width=6cm]{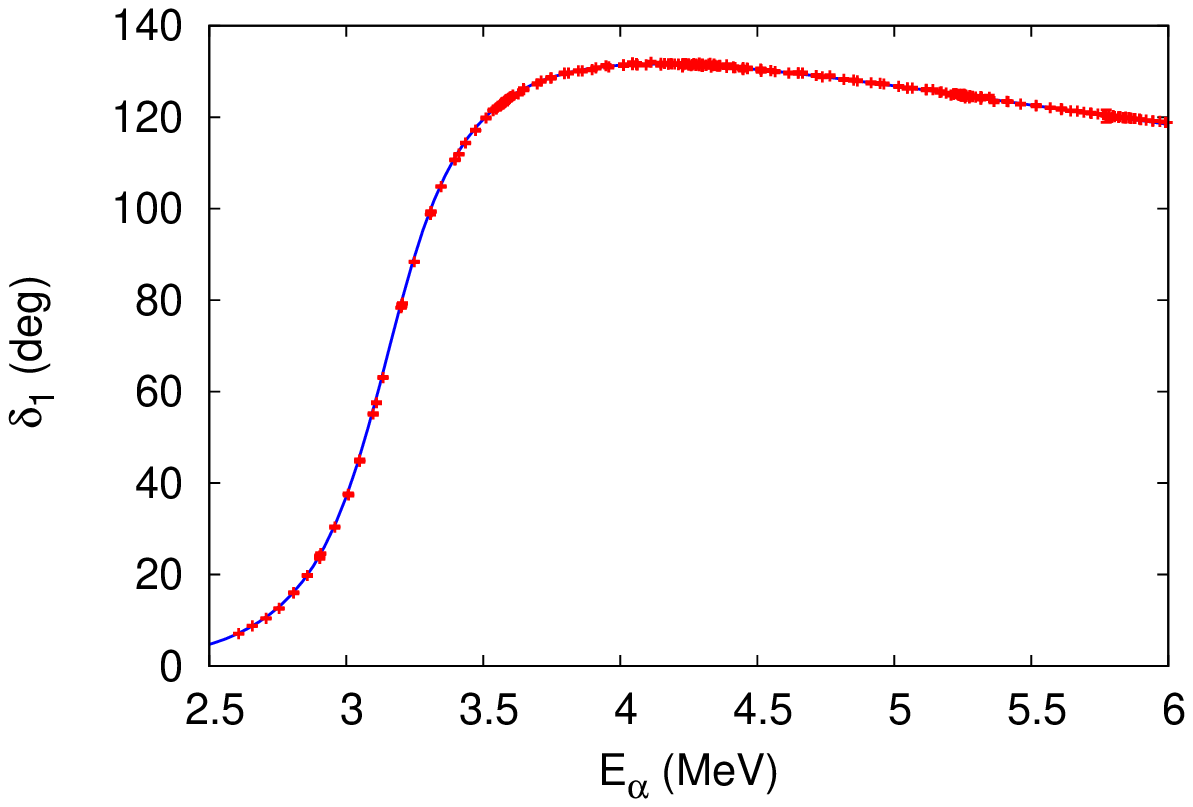}
\includegraphics[width=6cm]{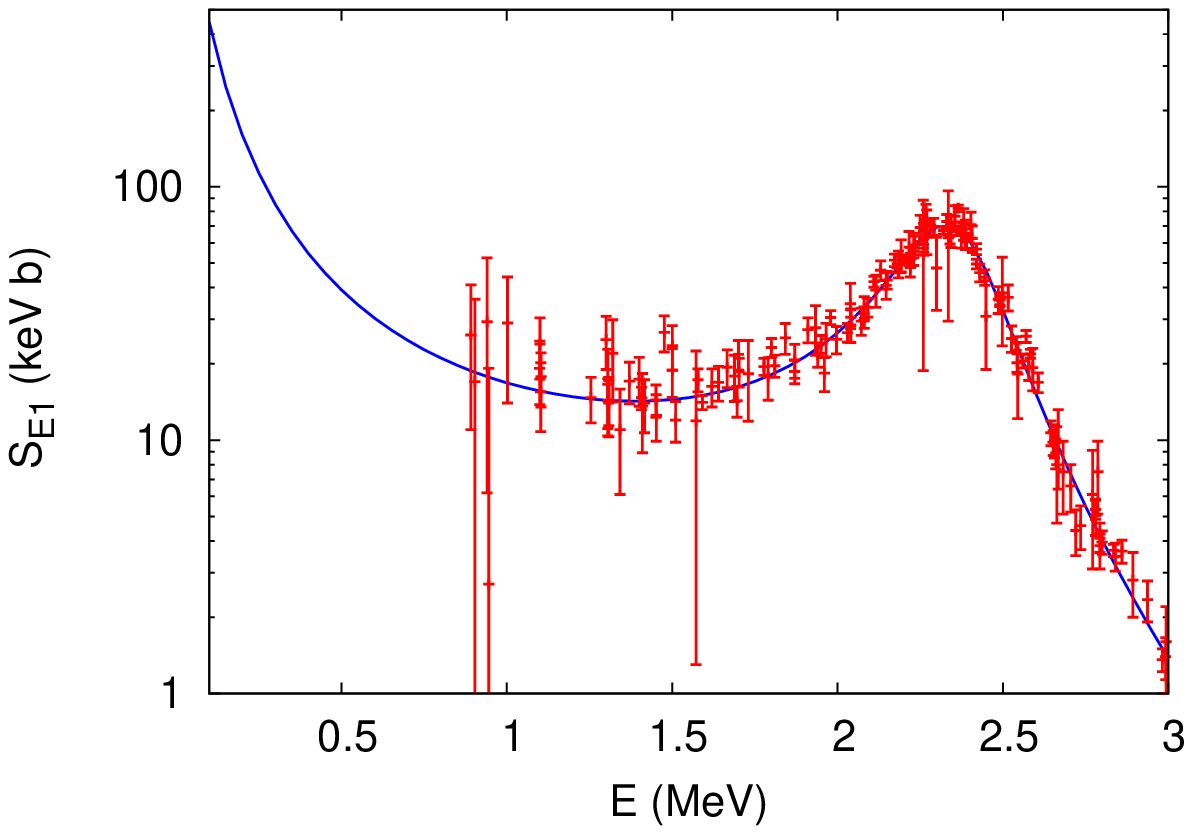}
\caption{
(Left panel) Phase shift, $\delta_1$, plotted by using the fitted effective
range parameters 
as a function of $E_\alpha$.
(Right Panel) $S_{E1}$ factor plotted by using the fitted parameters 
as a function of $E$.
The experimental data are also displayed in the figures.
\label{figs}
}
\end{center}
\end{figure}

In the left panel of Fig.~\ref{figs}, we show the data and the fitted curve
of the phase shift
and find that the fitted curve well reproduces the data. 
The remaining two parameters, $h^{(1)R}$ and $y^{(0)}$, in the amplitudes
are fitted to the $S_{E1}$ data~\cite{detal-17}, and we obtain
$h^{(1)R} = -6.95(11)\times 10^2$~MeV$^3$ and 
$y^{(0)} = 0.495(18)$~MeV$^{-1/2}$,
where the number of the data is $N=151$ and $\chi^2/N\simeq 1.72$.
In the right panel of Fig.~\ref{figs}, we show the data and the fitted 
curve for $S_{E1}$. 
At the Gamow peak energy, $E_G=0.3$~MeV, thus, 
we obtain $S_{E1}\simeq 58$~keV b.  
An error estimate of $S_{E1}$ is now under investigation.

\vskip 2mm

This work was supported by
the Basic Science Research Program through the National Research
Foundation of Korea funded by the Ministry of Education of Korea
Grant No. NRF-2016R1D1A1B03930122.

%

\end{document}